\documentclass[12pt, prd, showpacs]{revtex4}
\usepackage{amssymb}
\usepackage{amsmath}

\input{tcilatex}

\begin{document}

\title{Super-Penrose process due to collisions inside ergosphere}
\author{O. B. Zaslavskii}
\affiliation{Department of Physics and Technology, Kharkov V.N. Karazin National
University, 4 Svoboda Square, Kharkov 61022, Ukraine}
\affiliation{Institute of Mathematics and Mechanics, Kazan Federal University, 18
Kremlyovskaya St., Kazan 420008, Russia}
\email{zaslav@ukr.net }

\begin{abstract}
If two particles collide inside the ergosphere, the energy in the centre of
mass frame can be made unbound provided at least one of particles has a
large negative angular momentum (A. A. Grib and Yu. V. Pavlov, Europhys.
Lett. 101, 20004 (2013)). We show that the same condition can give rise to
unbounded Killing energy of debris at infinity, i.e. super-Penrose process.
Proximity of the point of collision to the black hole horizon is not
required.
\end{abstract}

\maketitle

\section{Introduction}

Investigation of high energy collisions in the black hole background now
attracts much attention. It was stimulated by the observation that collision
of two particles moving towards a black hole can produce an \ indefinitely
large energy $E_{c.m.}\,\ $in the centre of mass frame \cite{ban}. The same
happens if particles move in opposite directions \cite{pir1} - \cite{pir3}.
In this context, there are two different issues, connected with obtaining
(i) large energies in the centre of mass $E_{c.m.}$ and (ii) large Killing
energies $E$ of debris at infinity. It turned out that there are serious
restrictions on $E$ even in spite of large $E_{c.m.}$ since strong
gravitational redshift almost compensates the excess of energy \cite{p} - 
\cite{z}. Nonetheless, there exist scenarios, in which $E$ is also
significantly amplified or even unbounded (hence, extraction of energy from
a black hole is big) \cite{shnit} - \cite{mod}. Such cases are called the
super-Penrose process in \cite{card}, and we stick to this terminology.
However, the super-Penrose process near black holes has its own severe
restrictions \cite{pir15}, \cite{epl1}.

Up to now, all discussion in literature concerning the super-Penrose process
in the black hole background applied to collisions near the event horizon
only. In the present paper we show that there exists an alternative
mechanism in which large $E$ and $E_{c.m.}$ are compatible with each other.
It is based on the Grib-Pavlov mechanism of collision. It was shown in \cite%
{gpergo} that if two particles collide inside the ergosphere of the Kerr
metric and at least one of particles has the large negative angular momentum 
$L$, the resulting $E_{c.m.}$ is also large. Later on, it was shown in \cite%
{mergo} that this is a universal property of ergoregions of generic axially
symmetric rotating black holes. In both aforementioned papers, only the
properties of $E_{c.m.}$ were considered. Below, we will see that for such a
type of collision the super-Penrose mechanism is possible. It is worth
stressing that, although in the scenarios under discussion $L$ for initial
particles is supposed to be large, their Killing energies $E$ are finite.
The similar combination (large $L$ and modest $E$) occurs also in some other
scenarios of high energy collisions - say, in the vicinity of magnetized
black holes \cite{fr} - \cite{innem} .

\section{Basic formulas}

Let us consider the metric%
\begin{equation}
ds^{2}=-N^{2}dt^{2}+g_{\phi }(d\phi -\omega dt)^{2}+\frac{dr^{2}}{A}%
+g_{\theta }d\theta ^{2}.
\end{equation}

We assume that all metric coefficient do not depend on $t$ and $\phi $. This
gives rise to the conservation of the energy $E=-mu_{0}$ and angular
momentum $L=mu_{\phi }$. Here, $m$ is the particle's mass, $u^{\mu }=\frac{%
dx^{\mu }}{dt}$ is the four-velocity, $\tau $ is the proper time. In what
follows, we restrict ourselves by motion in the equatorial plane $\theta =%
\frac{\pi }{2}$. For such a motion, one can always redefine the radial
coordinate to achieve $N^{2}=A$. Then, equations of motion read%
\begin{equation}
m\frac{dt}{d\tau }=\frac{X}{N^{2}}\text{,}
\end{equation}%
\begin{equation}
X=E-\omega L\text{,}  \label{x}
\end{equation}%
\begin{equation}
m\frac{d\phi }{d\tau }=\frac{L}{g_{\phi }}+\frac{\omega X}{N^{2}}\text{,}
\end{equation}%
\begin{equation}
m\frac{dr}{d\tau }=\sigma Z\text{, }Z=\sqrt{X^{2}-N^{2}(\frac{L^{2}}{g_{\phi
}}+m^{2})}.  \label{z}
\end{equation}%
Here, $\sigma =\pm 1$ depending on the direction of motion.

We assume the forward-in time condition $\frac{dt}{d\tau }>0$, whence (for $%
N\neq 0$)%
\begin{equation}
X>0.  \label{ft}
\end{equation}

If two particles 1 and 2 collide to produce particles 3 and 4, the
conservation of energy and angular momentum gives us%
\begin{equation}
E_{1}+E_{2}=E_{3}+E_{4}\text{,}  \label{en}
\end{equation}%
\begin{equation}
L_{1}+L_{2}=L_{3}+L_{4}.  \label{ang}
\end{equation}

The conservation of the radial momentum reads%
\begin{equation}
\sigma _{1}Z_{1}+\sigma _{2}Z_{2}=\sigma _{3}Z_{3}+\sigma _{4}Z_{4}\text{.}
\label{zz}
\end{equation}

It is implied that masses of all particles are fixed. Say, one can take $%
m_{3}=m_{1}$, $m_{4}=m_{2}$ for the elastic collision or $m_{3}=m_{4}=0$ for
annihilation of two initial particles into gamma quanta. The quantities $%
E_{1}$, $E_{2}$, $L_{1}$ and $L_{2}$ are fixed. We can also fix, say, $%
L_{4}. $ Then, three equations (\ref{en}) - (\ref{zz}) determine three
unknowns $E_{3}$, $E_{4}$, $L_{3}$.

\section{Scenarios of collision}

We assume that particle 3 moves outward right after collision and escapes,
so $\sigma _{3}=+1$. By assumption, particle 2 has a large negative angular
momentum $L_{2}=-\left\vert L_{2}\right\vert $. In general, eq. (\ref{zz})
is quite cumbersome algebraically. As our goal is just to demonstrate the
existence of the super-Penrose process, we will make several
simplifications. We assume that 
\begin{equation}
L_{1}=L_{2}b,L_{4}=aL_{2},  \label{l1}
\end{equation}
where $b$ and $a$ are numbers. Then, the conservation of the angular
momentum entails that 
\begin{equation}
L_{3}=L_{2}(1+b-a).  \label{l3}
\end{equation}%
In general, this still leads to rather bulky algebraic expressions in (\ref%
{zz}). We restrict ourselves by the case $a=1+b$, so $L_{3}=0$. This is
quite sufficient for our purpose - to demonstrate the existence of the
super-Penrose process. We are interested in the scenario in which $E_{3}$ is
large and has the order $\left\vert L_{2}\right\vert $, so we put%
\begin{equation}
E_{3}=\left\vert L_{2}\right\vert y+O(1)\text{, }  \label{e3}
\end{equation}%
\begin{equation}
y>0\text{.}  \label{y}
\end{equation}%
Correspondingly, 
\begin{equation}
E_{4}=-\left\vert L_{2}\right\vert y+O(1)\text{,}
\end{equation}%
so collision must occur inside the ergosphere where negative energies are
allowed. Then, 
\begin{equation}
X_{1}\approx \omega b\left\vert L_{2}\right\vert ,  \label{x1}
\end{equation}%
\begin{equation}
X_{2}\approx \omega \left\vert L_{2}\right\vert ,  \label{x2}
\end{equation}%
\begin{equation}
X_{3}\approx \left\vert L_{2}\right\vert y\text{,}  \label{x3}
\end{equation}%
\begin{equation}
X_{4}\approx \left\vert L_{2}\right\vert [\omega (1+b)-y]\text{.}  \label{x4}
\end{equation}%
Here, condition (\ref{ft}) for particle 1 requires $b>0$. It is satisfied
automatically for particle 2. It is also satisfied for particle 3, provided (%
\ref{y}) holds true. For particle 4 it gives us%
\begin{equation}
\omega (1+b)-y>0\text{.}
\end{equation}

By substitution into (\ref{zz}), we have in the leading order in $L_{2}$ the
equation%
\begin{equation}
\Omega _{c}(\sigma _{2}+\sigma _{1}b))-y=\sigma _{4}\sqrt{[(y-\omega
_{c}(1+b)]^{2}+(1+b)^{2}(\Omega _{c}^{2}-\omega _{c}^{2})}\text{,}
\label{s4}
\end{equation}%
where subscript "c' means that the corresponding quantity is taken in the
point of collision,%
\begin{equation}
\Omega =\sqrt{\frac{g_{00}}{g_{\phi }}}\text{, }g_{00}=-N^{2}+g_{\phi
}\omega ^{2}\text{.}  \label{omom}
\end{equation}%
As inside the ergosphere $g_{00}>0$, the quantity $\Omega $ is real. This is
again the point where the properties of the ergoregion come into play.

If eq. (\ref{s4}) has a positive root $y$ and for this root the forward-in
time condition (\ref{ft}) is satisfied, the super-Penrose process does occur.

Taking the square of (\ref{s4}), one can find that%
\begin{equation}
b\Omega _{c}^{2}(\varepsilon -1)+y[(1+b)\omega _{c}-\Omega _{c}(\sigma
_{2}+b\sigma _{1})]=0\text{,}
\end{equation}%
where%
\begin{equation}
\varepsilon =\sigma _{1}\sigma _{2}\text{.}  \label{ep}
\end{equation}%
As we are interested in the existence of the root $y\neq 0$, we must take 
\begin{equation}
\varepsilon =-1.  \label{e1}
\end{equation}%
Then,%
\begin{equation}
y=\frac{2b\Omega _{c}^{2}}{(b+1)\omega _{c}+\Omega _{c}\sigma _{2}(b-1)}%
\text{,}  \label{res}
\end{equation}%
\begin{equation}
\frac{X_{4}}{\left\vert L_{2}\right\vert }=\omega _{c}(1+b)-y=\frac{V}{%
(b+1)\omega _{c}+\Omega _{c}\sigma _{2}(b-1)}\text{, }  \label{x4r}
\end{equation}%
where%
\begin{equation}
V=\omega _{c}^{2}(1+b)^{2}+\omega _{c}\Omega _{c}\sigma
_{2}(b^{2}-1)-2b\Omega _{c}^{2}.  \label{v1}
\end{equation}%
Taking into account that according to (\ref{omom}), $\omega >\Omega $, it is
seen that for any $b>0$, the denominator in (\ref{x4r}) is positive. For the
numerator we have the condition $V>0$. We can rewrite (\ref{v1}) as%
\begin{equation}
V=2b(\omega _{c}^{2}-\Omega _{c}^{2})+\omega _{c}[\omega
_{c}(1+b^{2})+\sigma _{2}\Omega _{c}(b^{2}-1)].  \label{v2}
\end{equation}

As $\omega >\Omega ,$ it is clear from (\ref{v2}) that indeed $V>0$ for all
values of parameters.

One should be careful about the sign of $\sigma _{4}$ to avoid fake roots
after taking the square. This sign must coincide with that of the left hand
side of (\ref{s4}). It is straightforward to check that%
\begin{equation}
sign\sigma _{4}=sign[\omega _{c}\sigma _{2}(1-b^{2})-\Omega _{c}(1+b^{2})]%
\text{.}
\end{equation}%
For example, for $\sigma _{2}=-1$ and $b<1$, we must take $\sigma _{4}=-1$.
However, if, say, $\sigma _{2}=-1$, $b>1$ and $\omega >\Omega \frac{b^{2}+1}{%
b^{2}-1}$, we have $\sigma _{4}=+1$.

After collision, particle 3 moves away from a black \ hole. In doing so,
there are no turning points for it. Indeed, for this particle $L_{3}=0$ and%
\begin{equation}
Z_{3}^{2}=E_{3}^{2}-\frac{N^{2}}{g_{\phi }}m_{3}^{2}\text{,}
\end{equation}%
where we took into account the mass term which was discarded before in (\ref%
{z}) as small correction. Here, $E_{3}^{2}=O(L_{2}^{2})$ is large, the
second term is finite, so indeed $Z>0$. As far as particle 4 is concerned,
it falls into a black hole, if $\sigma _{4}=-1$. If $\sigma _{4}=+1$, it
moves from a black hole to the turning point and bounces back. Particle 4
cannot escape to the asymptotically flat infinity since its energy is
negative.

Thus we succeed in the sense that the unbounded energy $E_{3}$ is obtained.
For this purpose, it is necessary in our scenario with $L_{3}=0$ that
particles 1 and 2 move in the opposite directions before collisions
according to (\ref{ep}), (\ref{e1}).

\section{Energy in the centre of mass}

To evaluate the energy of the centre of mass $E_{c.m.}$, one can use the
known formula (see, e.g. eq. 19 of \cite{mergo} in which one should put $%
\theta =const$ ). It is more convenient to apply it to the pair of particles
3 and 4 than to the original ones 1 and 2 since now $L_{3}=0$. Then, it
follows from the aforementioned formula that%
\begin{equation}
E_{c.m.}^{2}=\frac{X_{3}X_{4}-\sigma _{4}Z_{3}Z_{4}}{N^{2}}\text{.}
\end{equation}%
Taking into account (\ref{z}), (\ref{x3}) and (\ref{x4}) we obtain in the
main approximation

\begin{equation}
E_{c.m.}^{2}=\frac{L_{2}^{2}\mu }{N^{2}},
\end{equation}%
\begin{equation}
\mu =y\{[\omega _{c}(1+b)-y]-\sigma _{4}\sqrt{[(\omega
_{c}(1+b)-y]^{2}-(1+b)^{2}(\omega _{c}^{2}-\Omega _{c}^{2})}\}\text{.}
\end{equation}%
Obviously, $\mu >0$ for any sign of $\sigma _{4}$. When $L_{2}^{2}%
\rightarrow \infty $, the energy $E_{c.m.}\rightarrow \infty $ as well.

\section{Collisions near the boundary of ergosphere}

In the previous section, we mainly concentrated on the case when an escaping
particle 3 has the angular momentum $L_{3}=0$. We found that the scenario
with unbounded $E_{3}$ are possible, provided $\left\vert L_{2}\right\vert $
is large enough. One can ask, whether this value is singled out and what
changes if $L_{3}\neq 0$. Although, as is said above, formulas become in
general cumbersome, there is a situation when analysis can be carried out
analytically in a rather simple form. This is the case when collisions occur
in the vicinity of the boundary of the ergoregion (see below).

As before, we assume that all angular momenta are proportional to $L_{2}$.
However, now both the coefficients $a$ and $b$ introduced in the beginning
of Section III are free parameters. Then, instead of eqs. (\ref{x3}) and (%
\ref{x4}) we have%
\begin{equation}
X_{3}\approx \left\vert L_{2}\right\vert [y+\omega (1+b-a)]\text{,}
\label{3a}
\end{equation}%
\begin{equation}
X_{4}\approx \left\vert L_{2}\right\vert [\omega a-y].  \label{4a}
\end{equation}

Equations (\ref{x1}) and (\ref{x2}) are still valid. Using the expression (%
\ref{z}) and taking into account (\ref{l1}), (\ref{l3}) we can write the
conservation of radial momentum (\ref{zz}) in main approximation in the form%
\begin{equation}
\left( \sigma _{1}b+\sigma _{2}\right) \Omega =\sqrt{y^{2}+2y\omega
(1+b-a)+\Omega ^{2}(1+b-a)^{2}}+\sigma _{4}\sqrt{y^{2}-2y\omega
a+a^{2}\Omega ^{2}}\text{.}  \label{pr}
\end{equation}

Here, we put $\sigma _{3}=+1$ for the escaping particle, as before. All
quantities in (\ref{pr}) are taken in the point of collision. When $a=1+b$,
we return to (\ref{s4}).

It is sufficient to find at least one scenario with unbounded $E_{3}$. Let
us choose 
\begin{equation}
\sigma _{1}=1,\sigma _{2}=-1,b>1\text{, }0<a<1+b\text{.}  \label{ab}
\end{equation}

By definition, $g_{00}=0$ on the boundary of the ergoregion, so $\Omega =0$
according to (\ref{omom}). Let collision occur inside the ergoregion but
very close to its boundary. Then, $\Omega \rightarrow 0$. We expect the
existence of the solution of (\ref{pr}) in the form%
\begin{equation}
y\omega =\Omega ^{2}x\text{,}
\end{equation}%
where $x=O(1)$. Although $\Omega $ is small, we imply that $\Omega
^{2}\left\vert L_{2}\right\vert $ is still large enough to have $E_{3}$
large according to (\ref{e3}). Now we can neglect in (\ref{pr}) terms $y^{2}$
inside the radicals and obtain the equation%
\begin{equation}
F(b)=b-1=f(x)\equiv \sqrt{(1+b-a)[2x+(1+b-a)]}-\sqrt{-2xa+a^{2}}\text{,}
\label{F}
\end{equation}%
where we chose $\sigma _{4}=-1$. Then, $x\leq x_{\max }=\frac{a}{2}$ to
guarantee that the expression inside the second radical is nonnegative.

We want to show that the positive solution of this equation with $x=O(1)$
does exist. It is seen from (\ref{F}) that $f(0)=1+b-2a$. Thus $f(0)<F$.
Meanwhile, the function $f(x)$ is monotonically increasing. Therefore, if we
achieve $f(\frac{a}{2})>F$, it will mean that in some intermediate point $%
0<x<x_{\max }$ the curve $f(x)$ intersects the line of constant $F$, so the
solution exists. This condition is rendered as%
\begin{equation}
b-1<\sqrt{(1+b-a)(1+b)}\text{,}
\end{equation}%
whence%
\begin{equation}
b>\frac{a}{4-a}\text{.}
\end{equation}%
This is quite compatible with (\ref{ab}), so the solution does exist. For
instance, we can take $b=3$, $a=2$. Then, eq. (\ref{F}) has the form%
\begin{equation}
1=\sqrt{x+1}-\sqrt{1-x}
\end{equation}%
that has a solution $x=\frac{\sqrt{3}}{2}$.

\section{Summary and conclusions}

Thus we showed that there exist scenarios in which particles collide inside
the ergoregion in such a way that not only (i) their energy in the centre of
mass diverges, but also (ii) the Killing energy of one of particles escaping
to infinity is unbounded. This suggests a more easy way of extracting energy
since now (i) there is no problem with the redshift and time delay \cite{mc}%
, \cite{com}, (ii) there is no problem with fine-tuning typical of high
energy collisions near the horizon \cite{ban}. The restrictions of the
super-Penrose process indicated in \cite{pir15}, \cite{epl1} are also
irrelevant now. We analyzed in detail two situations: (a) escaping particle
3 has $L_{3}=0\,\ $and (b) collision occurs very closely to the boundary of
the ergoregion. Meanwhile, it is clear from derivation that collisions with
unbounded $E_{3}$ can occur everywhere inside the ergosphere. 

A separate interesting question that remained outside the scope of the
present work is the conditions under which particles with unbounded negative 
$L$ can occur inside the ergosphere. It was pointed out in \cite{gpergo}
that such values can be obtained as a result of preceding collision.
However, more thorough inspection showed that the situation is not so simple
since there are obstacles against such a scenario in that collisions of
particles with finite $E$ and $L$ cannot give rise to indefinitely large
negative $L$ (see Sec. VI of \cite{frac} for details). Therefore, other
mechanisms should be relevant here to achieve large negative $L$ (thermal
fluctuations, variable or chaotic electromagnetic fields, etc.). They are
model-dependent and need separate treatment. Meanwhile, the results of our
work have general model-independent character irrespective of the way the
initial state is prepared.

One can hope that the observation made in the present work can be of use for
investigation of the role of collisional Penrose process in astrophysics 
\cite{dark}.

\begin{acknowledgments}
This work was funded by the subsidy allocated to Kazan Federal University
for the state assignment in the sphere of scientific activities.
\end{acknowledgments}

\end{document}